\documentclass[fleqn,usenatbib]{mnras}
\usepackage{txfonts}

\usepackage[T1]{fontenc}
\usepackage{ae,aecompl}

\usepackage{color, soul, ulem,graphicx, amsbsy}

\newcommand{\msano}{{\rm M}_\odot~{\rm yr}^{-1}}
\def\hd{HD\,219134}
\newcommand{\bx}{\mathbf{x}} 
\newcommand{\bRbar}{\bar{\mathbf{R}}}
\newcommand{\bxp}{\dot{\mathbf{x}}}

\newcommand{\e}[1]{\times 10^{#1}}

\defcitealias{paper1}{Paper 1}

\title[The HD\,219134 multi-planet system II]{Characterisation of the HD\,219134 multi-planet system II. Stellar-wind sputtered exospheres in rocky planets b \& c}
\author[A. A. Vidotto, et al.]
{A. A. Vidotto$^{1}$,\thanks{E-mail:aline.vidotto@tcd.ie}
	H. Lichtenegger$^{2}$,
	L. Fossati$^{2}$,
	C. P. Folsom$^{3}$,
	B. E. Wood$^{4}$,
	\and
	J. Murthy$^{5}$,
	P. Petit$^{3}$,
	A. G. Sreejith$^{2}$,
		{G. Valyavin$^{6}$}
\\
\\
$^{1}$ School of Physics, Trinity College Dublin, the University of Dublin, Dublin-2, Ireland\\
$^{2}$ Space Research Institute, Austrian Academy of Sciences, Schmiedlstrasse 6, A-8042 Graz, Austria\\
$^{3}$ IRAP, Universit\'e de Toulouse, CNRS, UPS, CNES, 31400, Toulouse, France\\
$^{4}$ Naval Research Laboratory, Space Science Division, Washington, DC 20375, USA\\$^{5}$ Indian Institute of Astrophysics, Bangalore 560 034, India\\
$^{6}$ Special Astrophysical Observatory, Laboratory of Stellar Magnetism, Nizhnii Arkhyz, Karachai-Cherkessian Republic, 369167, Russia
}

\date{Accepted XXX. Received YYY; in original form ZZZ}

\pubyear{}

\begin{document}
\label{firstpage}
\pagerange{\pageref{firstpage}--\pageref{lastpage}}
\maketitle

\begin{abstract}
We present a 3D study of the formation of {refractory-rich} exospheres around the rocky planets HD219134b and c. These exospheres are formed by surface particles that have been sputtered by the wind of the host star. The stellar wind properties are derived from magnetohydrodynamic simulations, which are driven by observationally-derived stellar magnetic field maps, and constrained by Ly-$\alpha$ observations of wind mass-loss rates, making this one of the most well constrained model of winds of low-mass stars.
The proximity of the planets to their host star implies {a high flux of incident stellar wind particles, thus the sputtering process is sufficiently effective} to build up relatively dense, {refractory-rich} exospheres. The sputtering releases refractory elements from the entire dayside surfaces of the planets, with elements such as O and Mg creating an extended neutral exosphere with densities larger than 10~cm$^{-3}$, extending to several planetary radii. For planet `b', the column density of O{\sc i} along the line of sight reaches $10^{13}$~cm$^{-2}$, with the highest values found ahead of its orbital motion. This asymmetry would create asymmetric transit profiles. To assess its observability, we use a ray tracing technique to compute the expected transit depth of the O{\sc i} exosphere of planet `b'. We find that the transit depth in the O{\sc i} $1302.2$~\AA\ line is $0.042\%$, which is a small increase relative to the continuum transit ($0.036\%$). This implies that the sputtered exosphere of HD219134b is unlikely to be detectable with our current UV instruments.
\end{abstract}

%
\section{Introduction}\label{sec.intro}
Winds {from low mass, main-sequence stars} are formed from streams of {charged} particles that outflow from stars and, thus, permeate the interplanetary medium. As they make their way towards the interstellar medium, stellar wind particles drag along the stellar magnetic field. This magnetised plasma then interacts with orbiting exoplanets in a similar way as the solar wind interacts with solar system planets.

 The nature of the wind-planet interaction mainly depends on whether a planet is magnetised or not and whether it has a thick gaseous atmosphere or not \citep[e.g.][]{russell2016}. The solar system offers us some illustrations of different types of interactions. For example, the Earth's atmosphere is shielded from the direct interaction with the solar wind due to the presence of an extended (10--15\,$R_{\oplus}$) magnetosphere, which carves a cavity in the solar wind plasma, deflecting it around our magnetosphere. The magneotsphere of Mercury, which has a much weaker magnetic field, also carves a cavity in the solar wind, however with a significantly smaller size of only about 1.5 Mercury radii \citep{2013pss3.book..251B}. Mars, instead, is an example of a planet whose atmosphere directly interacts with the solar wind, creating an induced magnetosphere \citep{2011SSRv..162..113B}. 

In analogy to the interactions between the solar wind and the solar system planets, exoplanets {will also} interact with the winds of their host stars. Important differences can however exist, as both the architecture of known exoplanetary systems and the properties of the host stars can be significantly different from those of the solar system \citep{2015MNRAS.449.4117V}. For example, close-in exoplanets orbit at very short distances from their hosts, where the stellar wind is denser and the embedded  magnetic field is stronger, when compared to planets orbiting at large distances. For this reason, close-in planets usually interact with harsher stellar wind environments than farther out planets. Likewise, planet-hosting stars might be quite different from the Sun (e.g., more magnetically active), such that even planets that are not necessarily too close from their host stars might interact with winds of significantly different properties \citep[e.g.,][]{2013A&A...557A..67V}.

For planets similar to Mars (i.e., non-magnetised and with a thin, yet collisional atmosphere) or Mercury (i.e., 
weakly-magnetised and with a tiny, non-collisional atmosphere), the stellar wind interacts directly with either their atmosphere or  solid surface. Although lacking a substantial atmosphere, bodies like Mercury may hold a tenuous (i.e., non-collisional) gaseous envelope, forming their exospheres. This exosphere is made up of particles sputtered from the surface by precipitating solar wind protons and following ballistic orbits around the planet. Photoionisation of these neutral particles creates an ion population in addition to the ions directly ejected from the surface. 

The continuous supply of the exosphere is sustained by various surface-release processes, like photon stimulated desorption, thermal evaporation, micrometeoroid impact, and ion sputtering. Among these processes, sputtering is considered to be the most energetic mechanism, leading to particles with energies of up to several hundreds of eV, distinctly exceeding the escape energies of many species at the surface of Earth-like planets. For Mercury, the surface, exosphere, and magnetosphere, together with the solar wind, constitute a complex and strongly coupled system dominated by the interaction of the neutral and ionised particles with the surface and the magnetospheric plasma. The upcoming ESA mission {\it BepiColombo} to Mercury is specifically devoted to the investigation of this highly dynamic and complex hermean environment \citep{2005SSRv..117..397M, 2007SSRv..132..433K}.

In the case of close-in airless exoplanets, sputtering of their surfaces may be stronger than at Mercury, thus raising the question whether their expected exospheres might be observable. We study here the multi-planetary system \hd, which hosts six detected planets to date. Five of them orbit the star at separations smaller than $0.4$~au, while one distant gaseous giant planet orbits at $\approx 3$~au. The host star is a K3 dwarf, with an estimated age of $11.0\pm2.2$~Gyr \citep{2017NatAs...1E..56G} and an average large-scale surface magnetic field of $\approx 2.5$~G (\citealt{paper1}, henceforth \citetalias{paper1}). The two inner-most planets, \hd\,b and c, which are likely tidally locked, are {observed in transit} and present Earth-like densities \citep{2017NatAs...1E..56G}. {Their low-gravities suggest that both planets lost through escape their primary hydrogen-dominated atmospheres, presumably accreted during formation. The removal of their primary atmospheres is likely to have happened while the system was still young and the star was active \citep[e.g.,][]{2011A&A...532A...6S,2015A&A...577L...3T}.}

Following the escape of their primary atmospheres, most likely the solidification of the magma oceans led to the formation of steam CO$_2$-dominated atmospheres \citep{2008ApJ...685.1237E,2012AREPS..40..113E,2018A&ARv..26....2L}. Depending on the past evolution of the high-energy {(X-ray and extreme ultraviolet, collectively called XUV henceforth)} stellar radiation, the planets may have lost also the CO$_2$-dominated secondary atmospheres \citep{2009ApJ...703..905T}, leaving behind the bare planetary surfaces directly interacting with the stellar wind. 

In this work, we start from the assumption that both planets have lost their CO$_2$-dominated atmosphere and do not host a significant magnetic field. Under these assumptions, the close proximity of both planets to the host star, and thus high {proton flux} of the incident  wind, make the surface sputter, similarly to what occurs on Mercury. Some of the sputtered planetary particles would then ionise, forming mostly neutral and ionised Na, O, Si, and Fe atoms \citep[e.g.,][]{2009ApJ...703L.113S,2011ApJ...742L..19M,2015P&SS..115...90P,2016ApJ...828...80K}. The structure and velocity of the material escaping from the planet would then be controlled by the stellar wind properties, radiation pressure, and interplanetary magnetic field carried by the stellar wind.

Here, we use state-of-the-art models of stellar wind and wind-induced sputtering to investigate the effects that the wind of \hd\ has on building up  exospheres on the two inner-most planets, how the wind interacts with it and whether these atmospheres can be observed. This paper is organised as follows. In Section~\ref{sec.wind}, we present our stellar wind model, which uses surface magnetic field maps derived from the Zeeman-Doppler Imaging technique \citepalias{paper1}. Our wind model is constrained by the observed mass-loss rate derived from Ly-$\alpha$ observations presented in \citetalias{paper1} of this series. We use the results of our stellar wind model to quantify the density, size, and distribution of the planetary exospheres forming as a result of stellar wind sputtering (Section~\ref{sec.sputtering}). {In Section~\ref{sec.observability}, we use a ray tracing technique to predict the observability of the exosphere through transmission spectroscopy in O{\sc i} lines.} Our discussion and conclusions are presented in Section~\ref{sec.conclusions}.

%
\section{Stellar wind modelling}\label{sec.wind}
\begin{figure}
\includegraphics[width=\columnwidth]{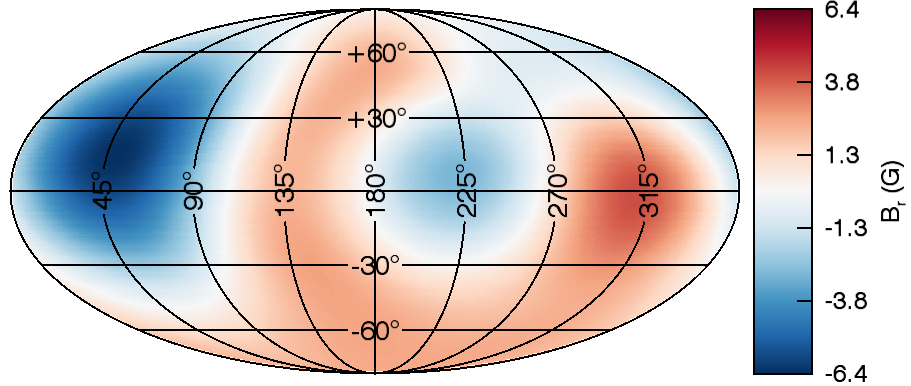}
\caption{Observationally-reconstructed surface magnetic field of the planet-hosting star \hd\ \citepalias{paper1}. The radial magnetic field component has an average absolute value of 1.9\,G.}
\label{fig.map}
\end{figure}

To model the stellar wind of \hd, we use the three-dimensional magnetohydrodynamics (MHD) code  BATS-R-US \citep{1999JCoPh.154..284P,2012JCoPh.231..870T}, which solves the set of ideal MHD equations in Cartesian coordinates, with adaptive-mesh refinement. The stellar wind model we use is the same as presented in, e.g., \citet{2015MNRAS.449.4117V,2017A&A...602A..39V}. For the stellar parameters, we take mass $M_\star = 0.81\,M_\odot$, radius $R_\star = 0.778\,R_\odot$ and rotation period of $42.2$~days \citepalias{paper1}. In our stellar wind model, the inner boundary conditions for the stellar magnetic field are taken to be the observationally-reconstructed surface magnetic field of \hd, \citepalias[Figure~\ref{fig.map}]{paper1}. The grid extends from $-20\,R_\star$ to $20\,R_\star$ in $X$, $Y$ and $Z$ directions, with the star placed at the centre of the grid and rotation axis aligned to positive $Z$-axis. The minimum and maximum cell sizes are $9.8\e{-3}$~$R_\star$ and $0.31$~$R_\star$, respectively. Closer to the star, the grid is better resolved, while further out, the resolution decreases (i.e., cell sizes are larger). The wind is polytropic, with a polytropic index of $1.05$, and consists of a fully ionised hydrogen plasma. The stellar wind total (electrons and protons) base density and temperature are set to $1.7\e{8}$~cm$^{-3}$ and 1.5\,MK, respectively. This set of base parameters was chosen such that the derived mass-loss rate from our models ($1.6\e{-14}$~$\msano$) matches the Ly-$\alpha$ astrospheric observations presented in \citetalias{paper1} ($\sim [0.5 - 2] \times 10^{-14}$~$\msano$). The simulation is evolved until it converges to a steady-state solution.

\begin{figure*}
\includegraphics[width=1.8\columnwidth]{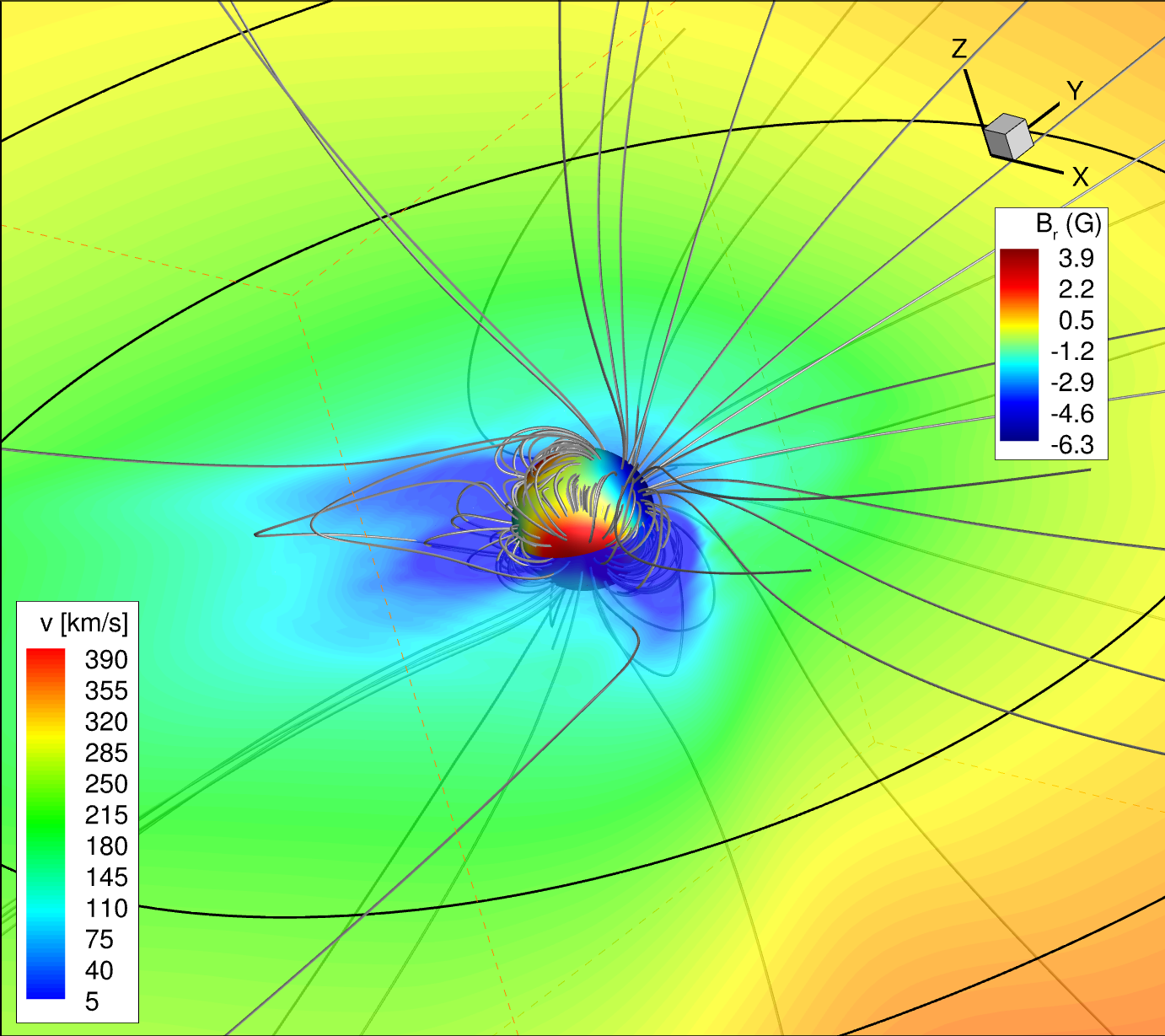}
\caption{Our stellar wind simulation uses the observationally-derived stellar magnetic field (colour scale on the right, \citetalias{paper1}) as the inner boundary condition for the stellar magnetic field. The stellar wind velocity is shown as a cut in the equatorial plane of the star (colour scale on the left) and the two circles indicate the orbital distances of planets `b' and `c'. The grey lines represent the stellar magnetic field, which is embedded in the stellar wind. The mass-loss rate derived in our simulations ($1.6\e{-14}~\msano$) is constrained by the Ly-$\alpha$ astrospheric observations presented in \citetalias{paper1}.}
\label{fig.wind3D}
\end{figure*}

Figure~\ref{fig.wind3D} shows the output of our simulations once reached steady-state. The colour scale on the right refers to the observed magnetic field map, while the colour scale on the left of the image refers to the stellar wind velocity, which is plotted in the equatorial plane of the star. The two circles indicate the orbital distances of planets `b' and `c', which we take to orbit in the equatorial plane of the star. The grey lines represent the stellar magnetic field, which is embedded in the stellar wind. The large-scale field of {the inner wind of \hd\ } resembles that of a tilted-dipole, with dipolar axis roughly parallel to the $Y$ direction. The wind velocity is faster when the magnetic field line has open topology at the surface of the star. Conversely, a slower wind appears above the closed-field lines, which are shaped as helmet streamers. This is similar to what is seen in eclipse observations of the solar corona at minimum. However, in the solar minimum case, the dipolar axis is essentially along the $Z$-axis and the wind structure is axisymmetric. This is also seen in simulations of axisymmetric dipolar fields \citep[e.g.][]{1971SoPh...18..258P,2009ApJ...699..441V}.

\begin{figure}
\includegraphics[width=0.98\columnwidth]{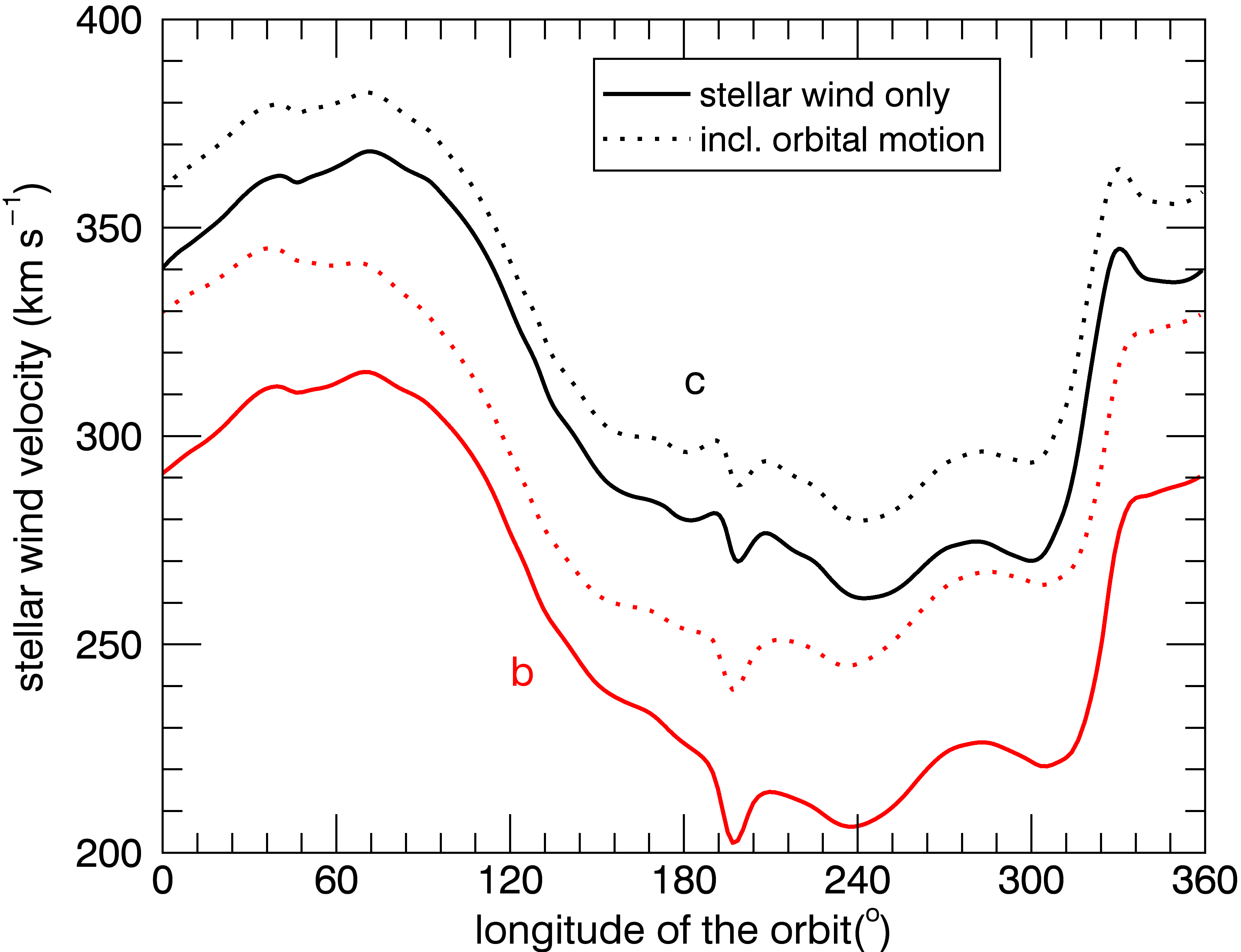}
\caption{Local stellar wind velocity around \hd\,b (red) and \hd\,c (black). Because planet `b' orbits at a closer distance to the star and the stellar wind accelerates with distance, the local velocity at planet `b' is smaller than at planet `c'.}
\label{fig.localwind}
\end{figure}

The fact that the magnetic field structure of \hd\ is non-axisymmetric creates variations on the local stellar wind conditions along the orbital path of the planets. The solid lines in Figure~\ref{fig.localwind} show the stellar wind velocity along the orbital path of \hd\,b (red) and \hd\,c (black). The wind velocity at planet `b' is smaller than that at planet `c', because the stellar wind accelerates with distance and planet `b' orbits at a closer distance to the star. Along one planetary year, both planets plunge through slow and fast winds. In the case of planet `b', the wind velocities vary from $200$ to $315$~km\,s$^{-1}$ in 3.09\,days. For planet `c', variations are from $260$ to $370$~km\,s$^{-1}$ along $6.76$~days. We remark that the velocity of the stellar wind particles that is seen by the planet is a vectorial sum of the wind velocity and of the planet's own orbital velocity \citep[e.g.,][]{2010ApJ...722L.168V}. The dashed lines in Figure~\ref{fig.localwind} show the magnitude of such a velocity, namely the velocity of the stellar wind in the reference frame of planets `b' and `c'.

Table~\ref{table1} summarises the local conditions of the stellar wind, averaged over one planetary orbit, at the location of planets `b' and `c'. Compared to Mercury ($0.39$~au from the Sun), the stellar wind proton densities are nearly two orders of magnitude higher than the local solar wind surrounding Mercury \citep[53\,cm$^{-3}$;][]{2013pss3.book..251B}. Likewise, the solar wind magnetic field surrounding Mercury \citep[40\,nT;][]{2013pss3.book..251B} is 13 and 5 times smaller than that surrounding \hd\,b and c, respectively. {The ram pressure of the incident solar wind protons at Mercury is $\sim 10^{-7}$~dyn~cm$^{-2} = 10^{-8}~$Pa, which is nearly 2 orders of magnitude smaller than the values around planets `b' and `c'}.

\begin{table}
\begin{center}
\caption{Physical properties of planets `b' and `c' (top) and of the local stellar wind (bottom), averaged over one planetary orbit.} \label{table1} 
\begin{tabular}{lcc}
\hline
\multicolumn{1}{r}{planet:} &  `b' &  `c'  \\
\hline\hline
\multicolumn{3}{l}{Planetary properties \citep{2017NatAs...1E..56G}:}\\ \hline
semi-major axis (au) & 0.038 & 0.065\\ 
semi-major axis $(R_\star)$ & 10.8 & 17.6 \\ 
orbital period (days) & 3.09 &  $6.76$ \\ 
 radius $R_p$ (R$_{\oplus}$) & $1.6$ & $1.5$ \\ 
 mass $M_p$ (R$_{\oplus}$) & $4.36$ & $4.34$ \\ 
average density (g~cm$^{-3}$) & $6.34$ & $6.95$  \\ 
\hline
\multicolumn{3}{l}{Stellar wind averaged local properties (this paper):} \\ \hline
 proton density ($10^3$~cm$^{-3}$) & $5.5$ & {$1.8$}  \\ 
 velocity (km~s$^{-1}$) & {$257$} &{ $312$}  \\ 
velocity incl.~orbital motion (km~s$^{-1}$) & {$290$} &{ $329$}  \\ 
{ram pressure of protons$^*$ ($10^{-6}$\,dyn~cm$^{-2}$)} & {$8$} &{ $3.2$} \\
 temperature ($10^{6}$ K) & {$1.10$} &{ $1.04$}  \\ 
 magnetic field  (nT) & {$530$} &{ $200$} \\ 
\hline
\end{tabular}
\end{center}
$^* 10^{-6}~{\rm dyn~cm}^{-2}= 10^{-3}~{\rm nanobar} = 10^{-7}~{\rm Pa}$
\end{table}

It is interesting to compare what would be the sizes of the magnetospheres of these rocky planets, in case they have a magnetic field similar to Earth's, i.e., a dipolar field with an equatorial strength of $B_p\sim 0.3$~G. To calculate the magnetospheric sizes, we equate the total pressure of the wind (the sum of thermal, magnetic, and ram pressures) to the magnetic pressure of the planet \citep[e.g.,][]{2015MNRAS.449.4117V}. Assuming the planets' magnetic field can be described as a dipole, the magnetospheric sizes would be $3.3$ and $4.0~R_p$ for planets `b' and `c', respectively. This is considerably smaller than the size of the Earth's magnetosphere, which is around 11\,$R_\oplus$, and is a consequence of the harsher environment surrounding close-in exoplanets \citep{2015MNRAS.449.4117V}. Interestingly, if these planets had a magnetic field similar to Mercury's \citep[$B_p\sim 1.95\e{-3}$~G;][]{2013pss3.book..251B}, their magnetic pressures would have been smaller than the total pressure of the local stellar wind, leading to magnetospheres essentially crushed onto the planetary surface. This would imply that the interaction between the stellar wind and the planet would take place very close to the planetary surface, thus sputtering would be significant even in the presence of a magnetic field.

%
\section{Stellar wind-induced sputtering in unmagnetised planets \hd\,b \& c}\label{sec.sputtering}
In this section, we investigate the effects that the wind of \hd\ has on building up an exosphere on planets \hd\,b and c.  
For the present study, we consider the two planets as airless rocky bodies and with a surface composition similar to that of Mercury. We further assume that the planets do not possess an intrinsic magnetic field so that the stellar wind can directly impact their entire dayside surfaces. As discussed in the end of Section~\ref{sec.wind}, Mercury-like magnetic fields would result in the same condition. The precipitating stellar wind protons are able to knock atoms off the surface with energies being sufficient for the sputtered atoms to form an extended tenuous atmosphere, i.e. an exosphere. 

\subsection{Description of the numerical model}

The numerical model used in this paper for calculating the three-dimensional exosphere densities is based on a modified version of the model used by \citet{2015P&SS..115...90P} to simulate the exospheric density of Mercury. 
For the sputter simulation,  we consider a steady-state situation, using stellar wind parameters averaged over one planetary orbit, as listed in Table~\ref{table1}, and thus we do not take into account the stellar wind variations displayed in Figure~\ref{fig.localwind}. {Our stellar wind model assumes a fully ionised hydrogen plasma. The presence of heavier species, such as He, can significantly contribute to the sputter yield, but these are not included in our stellar wind model.} We only study refractory elements that are ejected into the exosphere via stellar wind sputtering, and the release of volatile elements, like sodium or potassium, is not considered. Radiation pressure is not taken into account, as it does not markedly alter the trajectories of the sputtered particles. As we discuss further below, of all elements considered in this study, only calcium may be weakly influenced by the stellar radiation, leading to some additional small variations in its density distribution around the planets. 

\begin{table*}
\small
\caption{Elemental surface abundance in units of atom percent as modelled with the multiplicative composition modelling technique by \citet{2010P&SS...58.1599W} with a fixed Ca fraction of 1.67\%. The sputter yields, which refer to an incident angle of 45$^\circ$, and the cumulative photoionisation rates are shown for planets `b' and `c' for each element. Na and K are included in the surface composition model, but are not considered as sputtered elements.}
\label{tab.SurfaceComposition}
\begin{center}
\begin{tabular}{l|ccccccc} 
\hline
Species   & O & Na & Mg & Al & Si & P & S \\
\hline
Abundance (\%) &59.42&1.32&15.8&2.62&17.3&0.268&0.591 \\
Yield `b' (\%) &0.0600 & $ - $ & 0.0203 & 0.0012 & 0.0077  & 0.0001 & 0.0005 \\
Yield `c' (\%) &0.0658 & $ - $ & 0.0225 & 0.0018  & 0.0086 & 0.0003 & 0.0005 \\
Photoion. rate `b' (s$^{-1}$) & $3.91\times 10^{-5}$ & $ - $ & $6.72\times 10^{-5}$ & $1.05\times 10^{-1}$ &
 $2.81\times 10^{-3}$ & $1.61\times 10^{-4}$ & $2.48\times 10^{-4}$ \\
Photoion. rate  `c' (s$^{-1}$) & $1.34\times 10^{-5}$ & $ - $ & $2.3\times 10^{-5}$ & $3.6\times 10^{-2}$ &
 $9.59\times 10^{-4}$ & $5.49\times 10^{-5}$ & $8.48\times 10^{-5}$ \\
\hline\hline
Species   & K & Ca & Ti & Cr & Fe & Ni & Zn \\
\hline
Abundance (\%) & 0.030 & 1.670&0.014&0.041&0.611&0.004 &0.285\\
Yield `b' (\%) & $ - $ & 0.0017 & $<10^{-5}$ & 0.0001   & 0.0002  & $<10^{-5}$ & 0.0003 \\
Yield `c' (\%) & $ - $ & 0.0021 & $<10^{-4}$ & $<10^{-4}$ & 0.0003  & $<10^{-5}$ & 0.0004 \\
Photoion. rate `b' (s$^{-1}$) & $ - $ & $7.17\times 10^{-3}$ & $3.88\times 10^{-4}$ & $5.89\times 10^{-4}$ &
  $1.20\times 10^{-3}$ & $1.02\times 10^{-4}$ & $7.60\times 10^{-5}$ \\
Photoion. rate `c' (s$^{-1}$) & $ - $ & $2.45\times 10^{-3}$ & $1.32\times 10^{-4}$ & $2.01\times 10^{-4}$ &
 $4.09\times 10^{-4}$ & $3.47\times 10^{-5}$ & $2.60\times 10^{-5}$ \\
\hline
\end{tabular}
\end{center}
\end{table*}

For the elemental surface composition of the two planets, we assume abundances from the mineralogical model of \citet[][Table~\ref{tab.SurfaceComposition}]{2010P&SS...58.1599W}, which is based on available spectroscopic observations of Mercury's surface and which agrees reasonably well with {\it MESSENGER} results \citep{2015P&SS..115...90P}. The sputter yields for the various elements were calculated by means of the SRIM code \citep{Ziegler_online:2013}, where the {kinetic} energy of the impacting protons {$E_i$} is obtained from the stellar wind velocities in the frame of reference of the planet listed in Table~\ref{table1}. The equilibrium temperatures of \hd\,b and c are 1045 and 782\,K, respectively \citep{2017NatAs...1E..56G}. These temperatures are just below the minimum typical temperature of molten lava when it is first ejected from a volcanic vent. We therefore consider the surface of the two planets to be solid. This is further strengthened by the fact that we assume the planets to have no collisional atmosphere, which could have increased the surface temperature through greenhouse effect, as happens on Venus.

Since the angle between the incident ions and the surface normal varies with geographical location due to the sphericity of the planets, we calculated the sputter yields for different angles between 5$^\circ$ and 85$^\circ$, in
steps of 10$^\circ$. The yields for an angle of 45$^\circ$ are listed in Table~\ref{tab.SurfaceComposition}. The total production rate $p_i$ of the sputtered particles of species $i$ on the dayside is given by the particle flux $\Phi_{sw}$ of the stellar wind ions multiplied by the dayside surface area $A_{day}$ of the planet, by the corresponding sputter yield $Y_i$, and by a factor controlling the porosity $por$ of the regolith surface
\begin{equation}
p_i^{\rm b,c}=\Phi^{\rm b,c}_{sw}(\vartheta,\varphi)\,A^{\rm b,c}_{day}\,Y^{\rm b,c}_i(\vartheta,\varphi)\,(1-por)\,.
\end{equation}
Here, the superscripts $b$ and $c$ denote the two planets, while $\vartheta$ and $\varphi$ are the longitude and latitude of the precipitating protons, respectively. For the surface porosity, we assume a value of $por=0.3$ {for both planets} \citep{2005Icar..176..499C,2010P&SS...58.1599W}. 

Once the production rates of each species $p_i$ are known, sputtered particles are launched from random locations of the dayside surface with an initial kinetic energy distribution according to \citet{1969PhRv..187..768S} and \citet{2010P&SS...58.1599W}
\begin{equation}
\label{sigmund}
f(E_{\rm e}) = \frac{6E_b}{3-8\sqrt{E_{\rm b}/E_{\rm c}}}\,\frac{E_{\rm e}}{\left(E_{\rm e}+E_{\rm b}\right)^3}\,\left(1-\sqrt{\frac{E_{\rm e}+ E_{\rm b}}{E_{\rm c}}}\right)\\
\end{equation}
with
\begin{equation}
E_{\rm c}=E_{\rm i}\frac{4 m_1m_2}{(m_1+m_2)^2}\,,
\end{equation}
where $E_{\rm e}$ and $E_{\rm b}$ are, respectively, the kinetic energy and the surface binding energy of the sputtered particles with mass $m_2$, and $E_{\rm i}$ is the {kinetic} energy of the impacting {stellar wind proton} with mass $m_1$. {We remark that for Fe, the denominator of the first factor in Eq.~(\ref{sigmund}) becomes negative for $E_i\lesssim 450$ eV, while $E_i=436$ eV at the distance of planet `b'. To avoid this we set $E_i=450$ eV in the simulation for Fe sputtering at planet `b'.} The distribution of the angle $\theta$ between the surface normal and the initial trajectory of the sputtered particle is assumed to be proportional to $\cos{\theta}$, while the direction of the velocity vector in the horizontal plane is considered to be uniformly distributed between $[-\pi,\pi]$. 

Once sputtered from the planetary surface, a particle can be ionised by  stellar radiation. Assuming a transparent exosphere, the ionisation rate $f_i$ (1/s) of species $i$ can be obtained via \citep[e.g.,][]{2007GeoRL..34.1104M}
\begin{equation}
f_i =\int\limits_0^{\lambda_i^t}\sigma_i(\lambda)F_*(\lambda)d\lambda
\end{equation}
where $\lambda$ is the wavelength, $\sigma_i$ the photoionisation cross section, $\lambda_i^t$ is the photoionisation threshold for the species $i$ and $F_*(\lambda)$ the stellar spectral irradiance (photons~cm$^{-2}$~s$^{-1}$~nm$^{-1}$) derived in \citetalias{paper1}. The calculated photoionisation rates, based on the photoionisation cross sections available at {\tt http://phidrates.space.swri.edu/}, are listed in Table~\ref{tab.SurfaceComposition}. 

The forces applied on a sputtered atom are assumed to be the gravitational attraction of the planet, from which it is launched, and of the host star. Other possible forces, such as acceleration by stellar radiation, are considered to be small and are therefore neglected. Among the sputtered elements considered in this study, Ca is the most sensitive to radiation pressure, however, the radiative acceleration is still very small compared to the planet's gravitational acceleration.

The tracing of a particle is terminated either when it crosses the upper boundary at $6R_p$, or when it falls back onto the planetary surface, at which point we make the reasonable assumption of perfect sticking, which is valid particularly for refractory elements \citep{2010P&SS...58.1599W}. {If the particle is ionised along its trajectory, we still follow it, but change its statistical weight according to the ionisation probability.}

\begin{figure}
\includegraphics[width=\columnwidth]{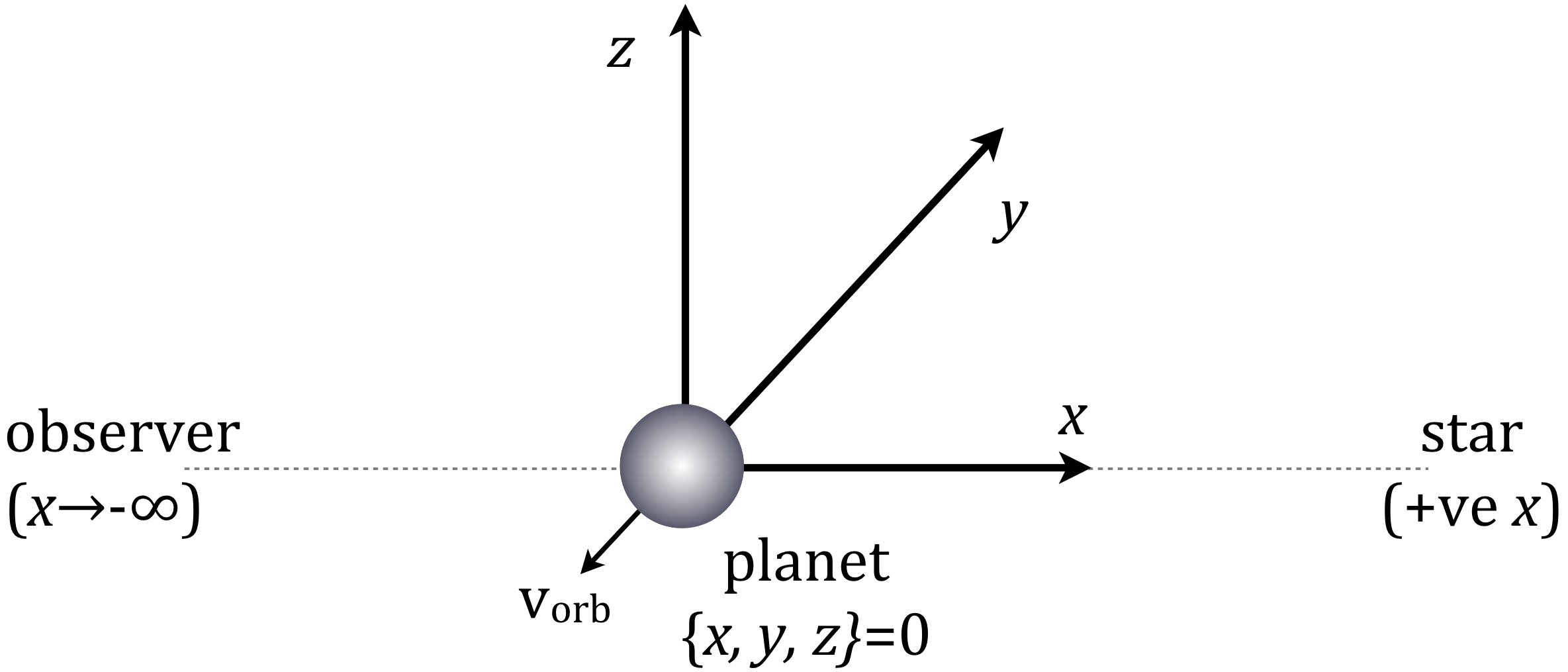}
\caption{Sketch of the coordinate system $K$: the planet is considered to be at the centre of the grid, with positive $x$ pointing towards the star and negative $x$ towards the observer. The orbital motion is towards negative $y$. }
\label{fig.frame}
\end{figure}

We introduce an inertial coordinate system $\bar{K}$  $(\bar{\mathbf{x}}=\{\bar{x},\bar{y},\bar{z}\})$ with its origin in the star-planet centre of mass and oriented in such a way that the planet's orbital plane coincides with the $\bar{x}\bar{y}$-plane. The position of the star and the planet are given by the vectors $\bRbar^*$ and $\bRbar$, respectively. The particles' motion is determined with respect to a coordinate system $K$ $(\bx=\{x,y,z\})$, whose origin is at the planetary centre, with the positive $x$-axis pointing towards the star, the positive $y$-axis being opposite to the direction of planetary motion and the $z$-axis is parallel to $\bar{z}$. Since we assume circular orbits of the planets and tidal locking, the $K$-frame rotates uniformly with an orbital frequency $\omega$, where the vector $\boldsymbol{\omega}$ is parallel to both the $z$- and $\bar{z}$-axes. A sketch of the $K$ reference frame is shown in Figure \ref{fig.frame}. The acceleration of a particle in the coordinate system $K$ is given by
\begin{equation}
\ddot{\mathbf{x}}=    -\frac{GM_p}{|\bx|^3}\bx-\frac{GM_\star}{|\bx^*|^3}\bx^*    -\boldsymbol{\omega}\times(\boldsymbol{\omega}\times\bRbar)-2\boldsymbol{\omega}\times\bxp-\boldsymbol{\omega}\times(\boldsymbol{\omega}\times\bx),
\end{equation}
where 
\begin{equation}
\bx^*=\bRbar-\bRbar^*+\bx \, ,
\end{equation}
\begin{equation}
\bar{\mathbf{R}} =[R\cos(\omega t+\pi),R\sin(\omega t+\pi),0] \, ,
\end{equation}
\begin{equation}
\bRbar^* = [R^*\cos\omega t,R^*\sin\omega t,0] \, .
\end{equation}
Here, $M_p$ and $M_\star$ are the planetary and stellar masses, respectively, $G$ is the gravitational constant, $R=|\mathbf{R}|$ and $R^*=|\mathbf{R}^*|$. Our 3D grid extends from $-6R_p$ to $6R_p$ in each axis and each axis is divided in 201 elements. This gives a linear resolution of $0.0597R_p$ for each cell in our grid.

\subsection{Exospheric densities}

\begin{figure*}
\includegraphics[width=2\columnwidth]{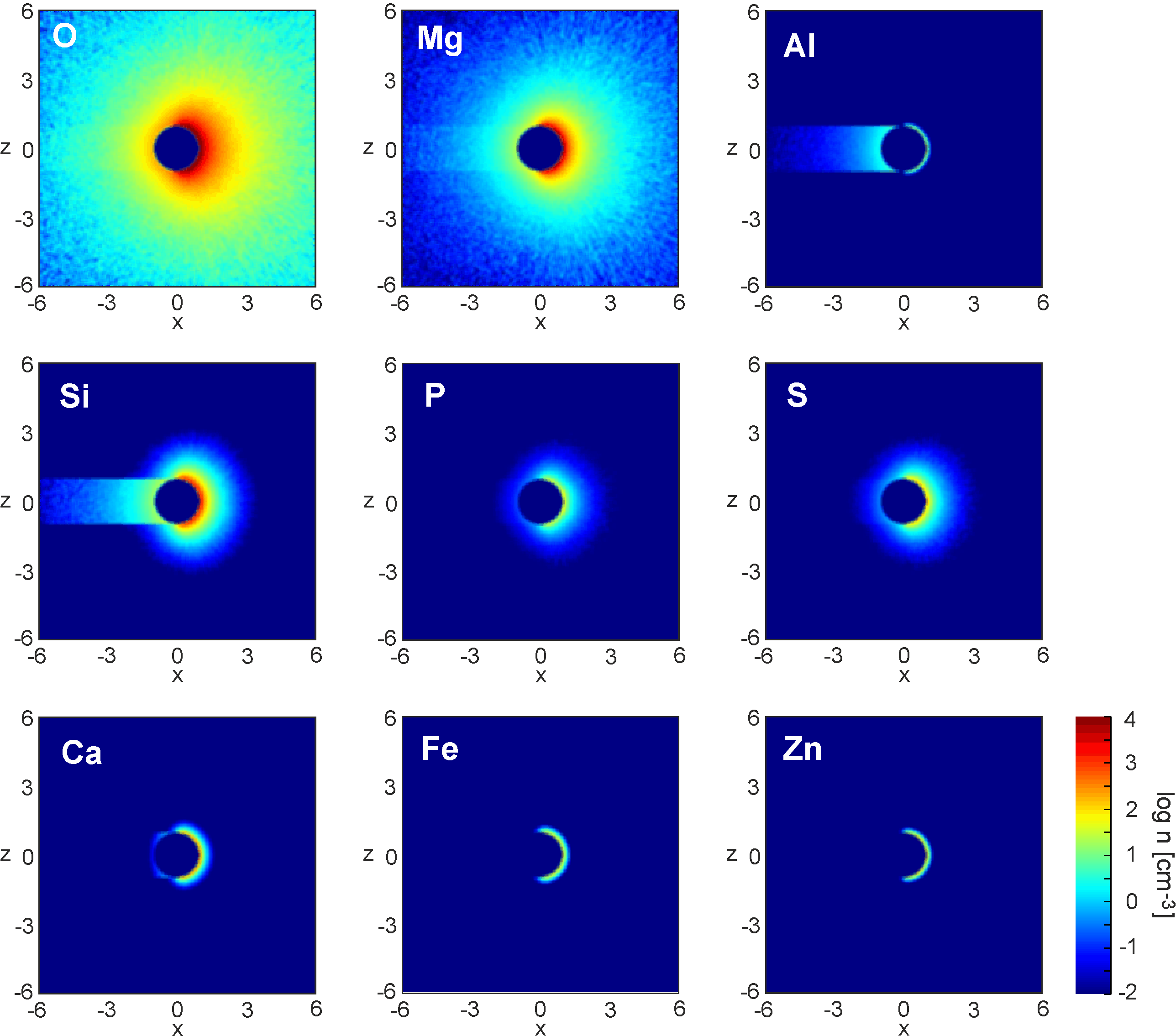}
\caption{Number density for sputtered elements in the noon-midnight plane in the vicinity of planet `b'. The $x$-axis is in the planet-star direction, with the star on the right and the $z$-axis is perpendicular to the orbital plane. The unit of the axes is given in planetary radii. Here, centrifugal force, stellar gravity and photoionisation are included. Appendix \ref{ap.physical} shows the effects of each of these physical ingredients in our models.}
\label{fig.densities.b}
\end{figure*}

Figure \ref{fig.densities.b} displays the number density of the  modelled sputtered refractory elements for planet `b'. In general, the exospheres are less extended and denser compared to that of Mercury. This is mainly due to the stronger gravitational field of \hd\,b and c and to the higher photoionisation rate caused by their close distance to the star. The exosphere is more extended in O, Mg, Si, P and Ca, respectively. 
{In our simulations, we assume the star as a point source and assume} no photoionisation in the geometrical shadow of the planet, which leads to a density enhancement of neutral particles in the planetary shadow, as can be more easily seen in the distribution of Al density.  
This enhancement is particularly distinct for those species that are easily ionised, such as aluminium, though in reality this transition will be less sharp and the density in the shadow will be somewhat reduced due to ionisation by scattered photons. {Moreover, due to the close distance of the planets, the extent of the star will also lead to the presence of a penumbra which might further modify the shape of the shadow. Therefore, a more realistic treatment of the stellar extension (instead of a point source) would make the planet shadow slightly more conical, but will have no effect on the observability of the exosphere.} The low iron density is not a result of strong ionisation, but it is due to the fact that the kinetic  energy distribution of sputtered Fe rapidly decreases above $\sim 10$~eV.
Since this is small compared to the iron escape energy of $\sim100$~eV, the atoms cannot reach high altitudes, confining them close to the surface. Iron atoms also remain neutral due to the short amount of time they spend in the exosphere.

\begin{figure*}\hspace*{-1.5mm}
\includegraphics[width=2\columnwidth]{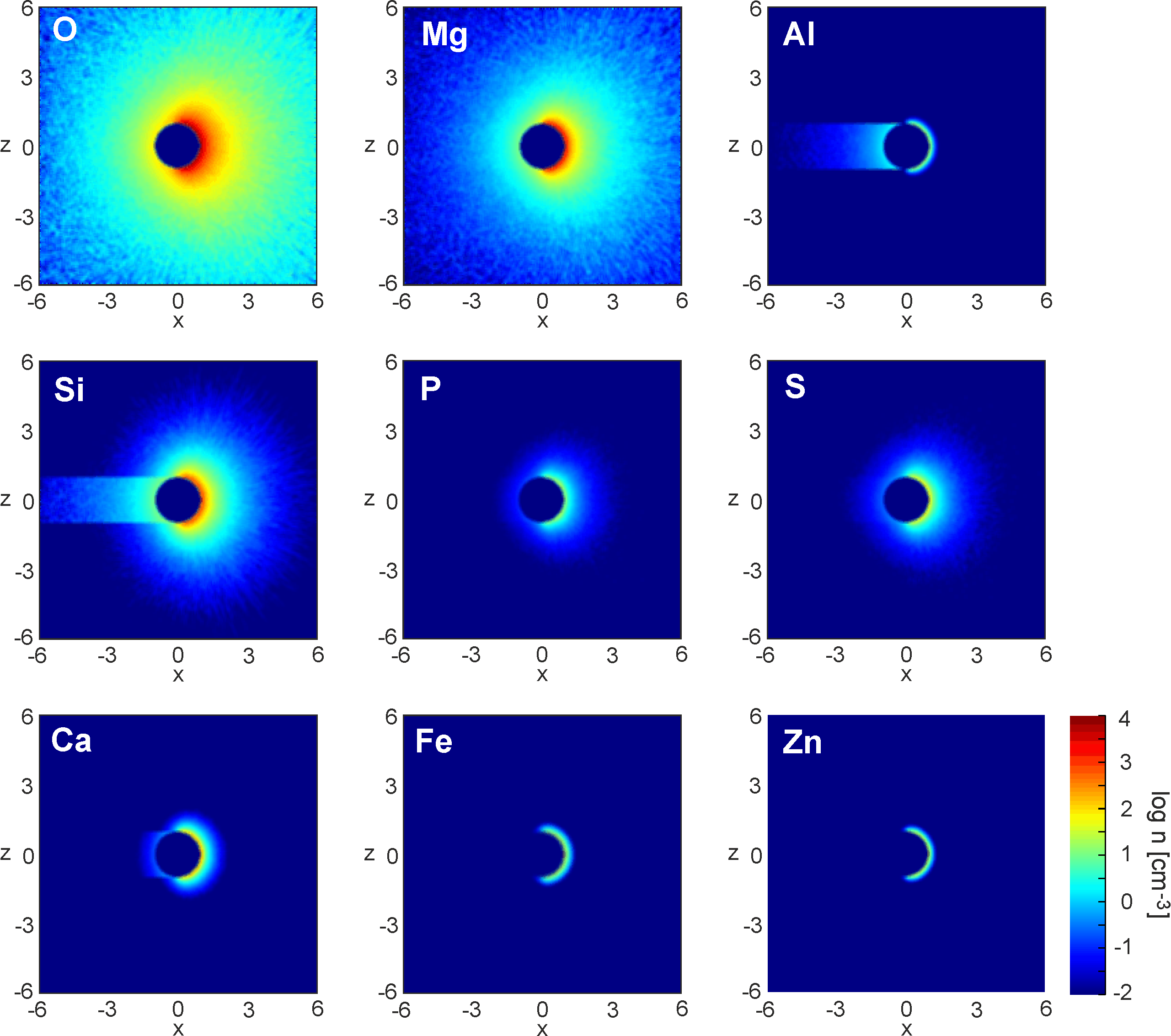}
\caption{Similar to Figure \ref{fig.densities.b}, but for for planet `c'.}
\label{fig.densities.c}
\end{figure*}

Figure~\ref{fig.densities.c} shows the number densities of the sputtered elements in the noon-midnight plane for planet `c'. The differences with respect to Figure~\ref{fig.densities.b} are mainly due to the weaker photoionisation rates and the different aberration angle of the stellar wind.

\begin{figure*}
\includegraphics[width=2\columnwidth]{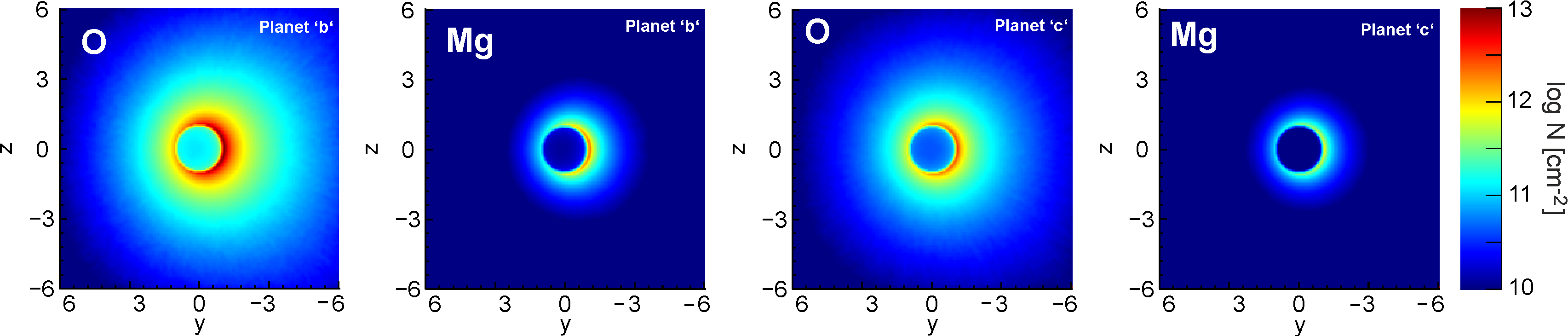}
\caption{Column densities for O and Mg at planet `b' (left panels) and planet `c' (right panels), when looking along the $x$-axis towards the star. Note the reversed orientation of the $y$ axis. The orbital velocity, pointing towards negative $y$, leads to an enhancement in the column density ahead of planetary motion  (i.e., to the right on the images above).}
\label{fig.coldensities}
\end{figure*}

Figure~\ref{fig.coldensities} shows the column densities for O and Mg obtained through our modelling for both planets when looking along the ${x}$-axis towards the star (i.e., in the plane of the sky). The column densities are obtained by integrating the number densities (Figures \ref{fig.densities.b} and \ref{fig.densities.c}) from $-6R_p$ to $6R_p$ along the $x$-axis. An interesting feature is the higher column densities ahead of the orbits of the planets (negative ${y}$-axis). This occurs due to the angle that the stellar wind makes with the day-side of the planet. The velocity of the stellar wind in the reference frame of the planet (i.e., taking into account its orbital motion) reaches planet `b' (`c') at an angle of 27.7 (18.7) degrees with respect to the planet-star line (positive ${x}$-axis).\footnote{For comparison, for Mercury, this angle is just a few degrees, because Mercury's orbital velocity is considerably smaller than the local solar wind velocity, such that the solar wind reaches Mercury roughly at its  day-side. If we were to calculate the column densities of Mercury's atmosphere in a similar way as done for  \hd\,b and c (Figure~\ref{fig.coldensities}), its distribution would  essentially be spherical in the $yz$ plane and centred in the planet.}  The misalignment between the velocity vector and the $x$ direction causes more sputtering ahead of the orbits of the planet, giving rise to the higher column densities there (negative ${y}$-axis). This asymmetry is more pronounced and more extended in the case of O in planet `b', with column densities reaching values of $\sim10^{13}$~cm$^{-2}$. If detectable, these higher densities indicate that planetary transits in, e.g., O or Mg lines, would be asymmetric, showing a longer ingress phase than the egress one. We discuss its observability in the next section.

%
\section{Observability of the exosphere of \hd\,b}\label{sec.observability}
Our results indicate that oxygen is the species that has the largest density, extends farther from the planetary surface, and remains in large part neutral. Here, we study the detectability of the oxygen corona surrounding \hd\,b. Because of the low gas density, the exosphere is best detectable at ultraviolet (UV) wavelengths and in particular in the far-UV (FUV). The FUV spectra of solar-like stars present three strong emission lines of neutral oxygen, making them ideal for the study of the detectability of the O  exosphere of \hd\, b. These three lines lie at $\lambda \approx 1302.2$, $1304.9$, and $1306.0$~\AA. Of these features, the one at shorter wavelength is the only resonance line {(i.e., from ground state)} and thus it is strongly affected by interstellar medium (ISM) absorption. Because the radial velocity of the star ($-18.8$~km~s$^{-1}$) is significantly shifted from the ISM absorption ($+7.3$~km~s$^{-1}$, \citetalias{paper1}), the ISM absorption probably do not obscure the expected exospheric O{\sc i} $1302$~\AA\ signature. 

\begin{figure*}
\includegraphics[width=0.325\textwidth]{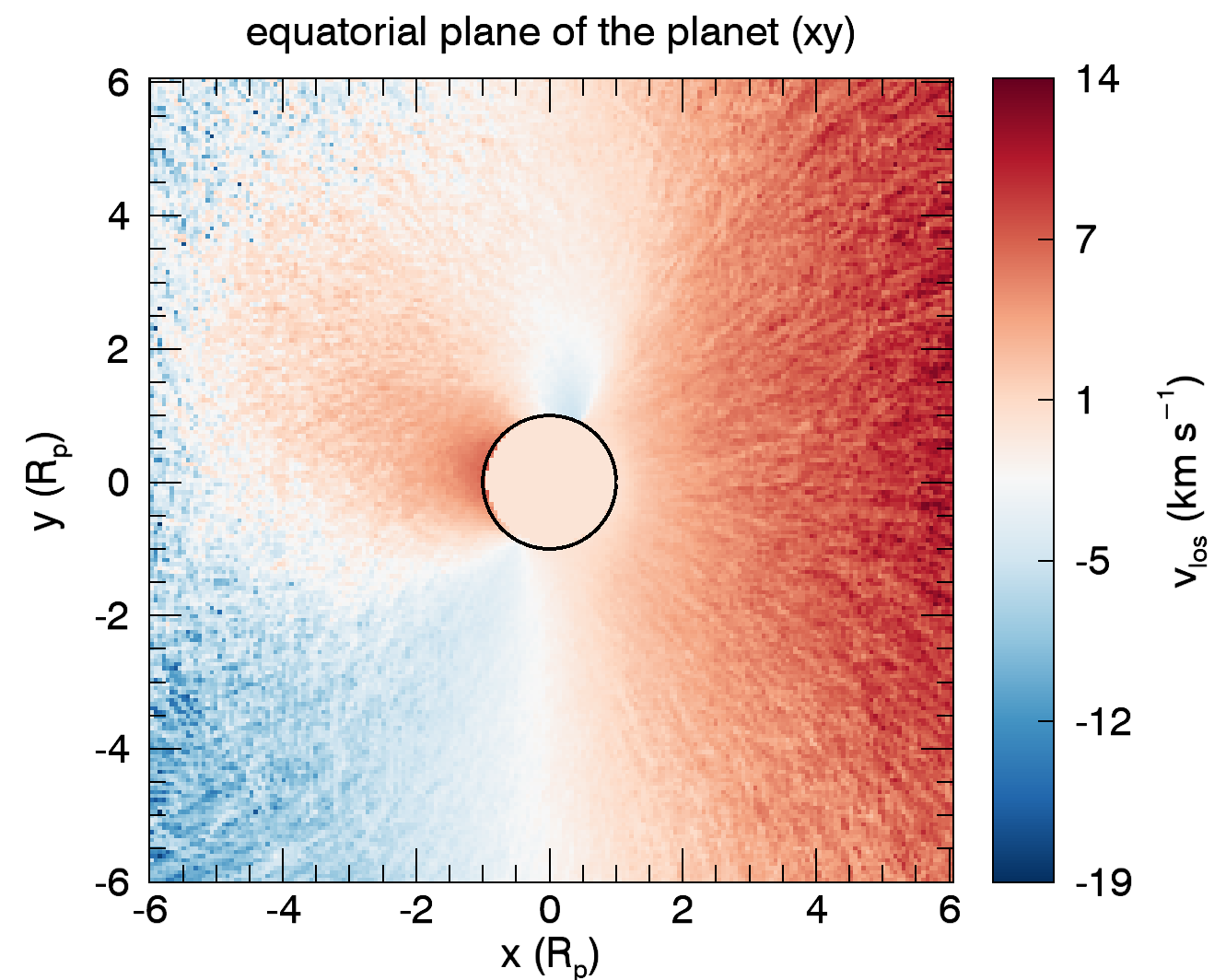}
\includegraphics[width=0.325\textwidth]{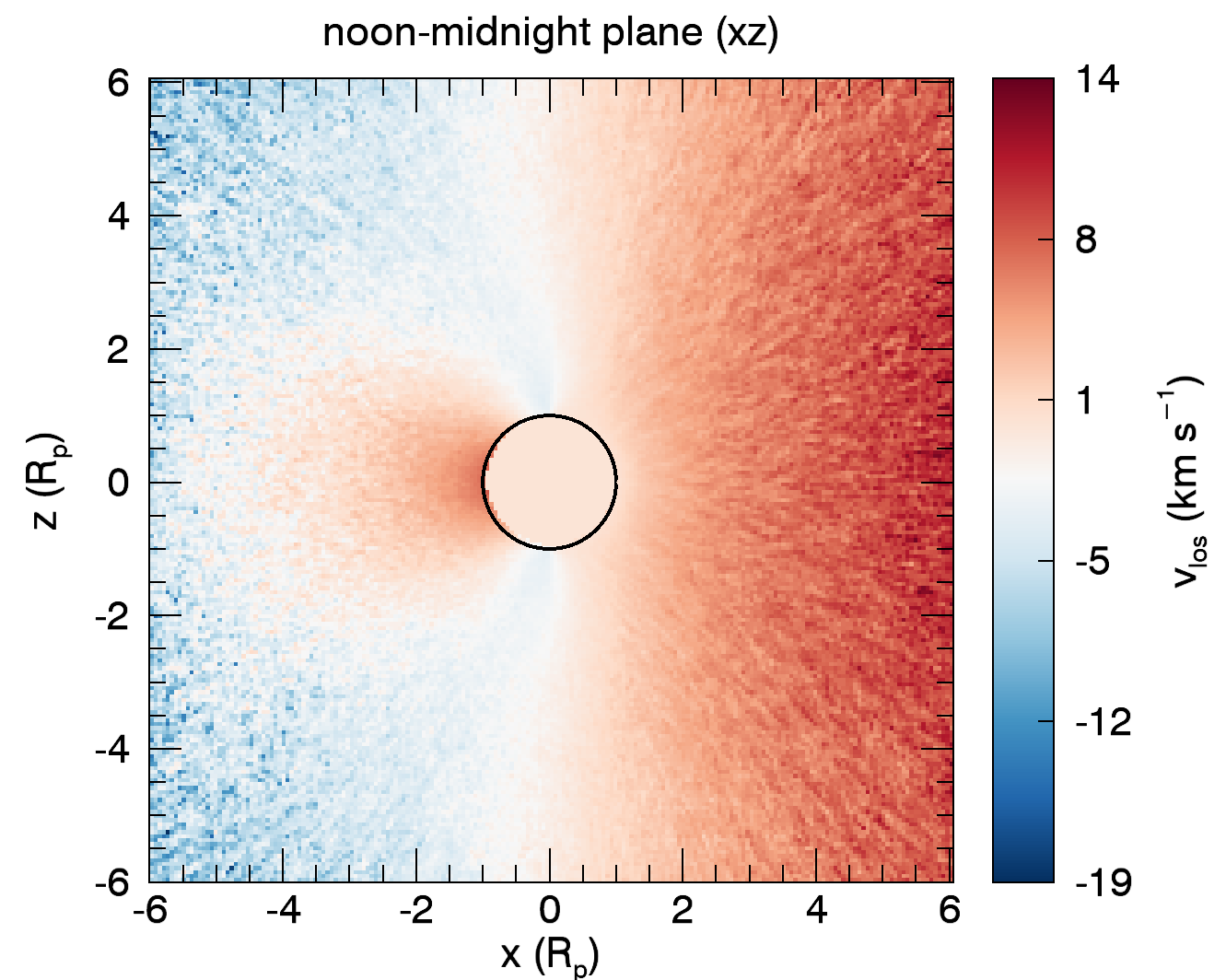}
\includegraphics[width=0.325\textwidth]{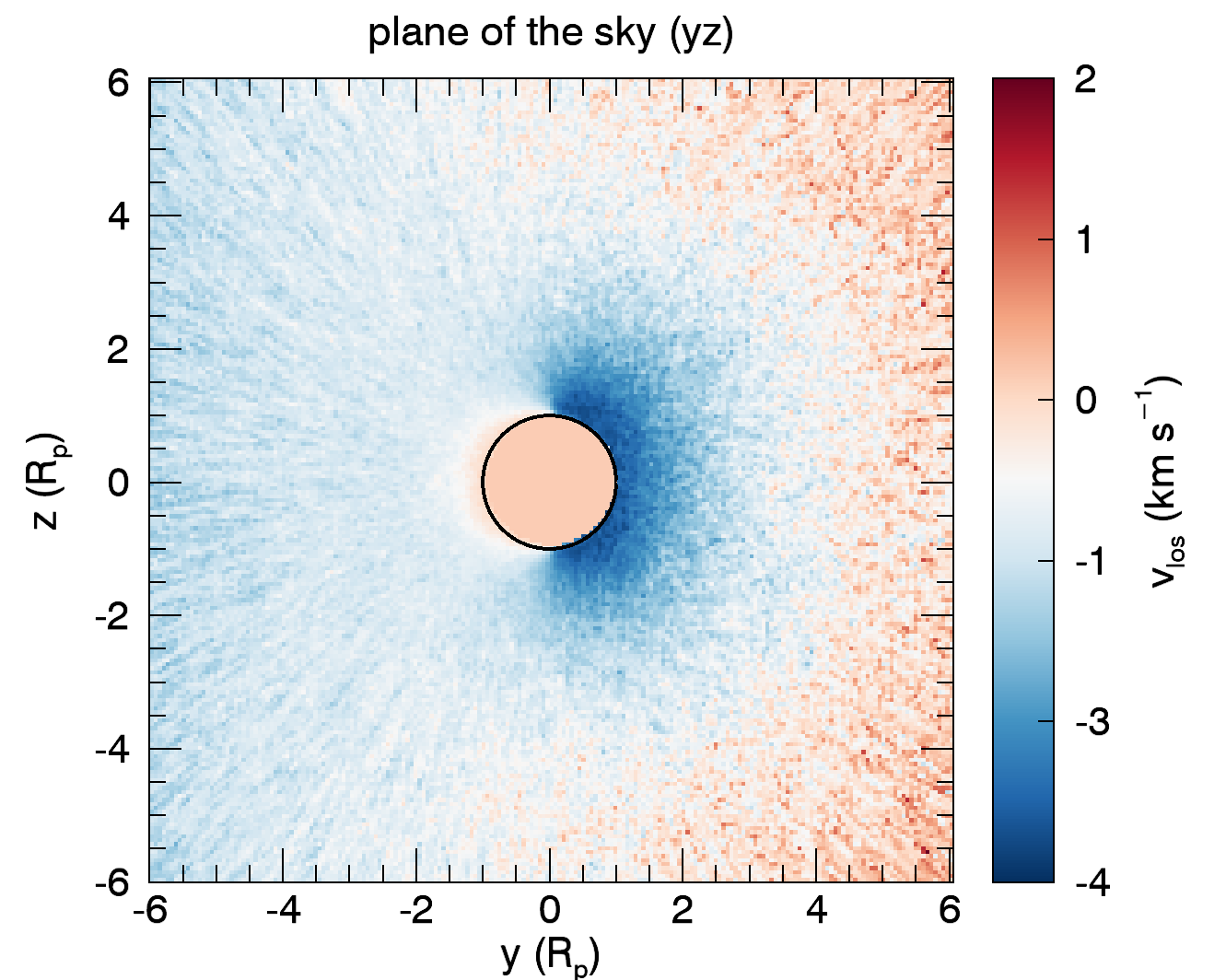}
\caption{Line-of-sight velocity of the oxygen exosphere of \hd\, b shown in cuts along the equatorial plane of the planet (left), the noon-midnight plane (middle) and the plane of the sky (right).}
\label{fig.Vlos}
\end{figure*}

Here, we calculate the transit depth in the resonance line O{\sc i} $1302.2$~\AA\ line using a ray tracing technique (see also \citealt{2014MNRAS.438.1654V,2018arXiv180602259V, 2018ApJ...855L..11O}). The optical depth as a function of the velocity $v$ (or wavelength) along one ray in  the direction that connects the observer to the star-planet system ($x$ direction) is 
\begin{equation}
\tau_v = \int n_{O} \sigma_0 \phi_v dx \, ,
\end{equation}
where $n_{O}$ is the number density of neutral oxygen in the exosphere of \hd\,b that is in the lower energy state of the $1302.2$~\AA\  transition, $\sigma_0$ is the O{\sc i} cross-section at line centre and $\phi_v$ is the line profile assuming Doppler broadening\footnote{The same calculation was done assuming a combined Doppler and Lorentz broadening (Voigt profile) and the transit depths obtained were the same as in the case of only considering Doppler broadening.}
\begin{equation}
\phi_v =  \frac{\lambda_0}{\sqrt{2  \pi} v_{\rm th} } \exp{-\left( \frac{v-v_{los}}{\sqrt{2}v_{\rm th}}\right)^2} \, .
\end{equation}
Here, the line centre is $\lambda_0=1302.168$~\AA , the thermal velocity is $v_{\rm th}=\sqrt{k_B T/ m_O}$, with $k_B$ the Boltzmann constant, and $m_O$ the mass of atomic oxygen, and the velocity offset from the line centre is $v-v_{los}$, with   $v_{los}$ being the velocity of the escaping atmosphere projected along the line-of-sight, i.e., in the $x$ direction (Figure \ref{fig.Vlos}).  We assume the temperature $T$ of the exosphere to be the equilibrium temperature of \hd\,b, which is $T=1045$~K. For the cross-section at line centre, we use $\sigma_0=\pi e^2 f/(m_e c) = 1.38\e{-3}$~cm$^{2}$~Hz, where the oscillator strength for the $1302.2$~\AA\ transition is $f=0.052$ \citep{1991JPhB...24.3943H} and was taken from the NIST database \citep{NIST_ASD}\footnote{\url{http://www.nist.gov/pml/atomic-spectra-database}}. We also assume that all the neutral oxygen in the exosphere of \hd\,b is in the ground state, which is a reasonable assumption for a low-density limit, in the absence of radiative pumping.   

During transit, the stellar emission along one ray is attenuated by $I_v/I_\star = e^{-\tau_v}$, where $I_\star$ is the specific intensity of the star and $I_v$ is the velocity-dependent specific intensity attenuated by the absorption from the planet and its atmosphere. We assume that the stellar disc emits a uniform specific intensity $I_\star$ at a given wavelength, neglecting centre-to-limb variation. To simulate an unresolved observation (point-source), we sum  the absorbed intensity $(1-e^{-\tau_v})$ per ray for all the  rays. The transit depth is then calculated as
\begin{equation}
\Delta F = \frac{\int\int (1-e^{-\tau_v})dy dz}{\pi R_\star^2} \, ,
\end{equation}
where $dy$ and $dz$ is the element of area associate to each ray in our simulation. Our grid contains 201 elements in each direction, which extends from $-6R_p$ to $6R_p$. 

\begin{figure}
\includegraphics[width=\columnwidth]{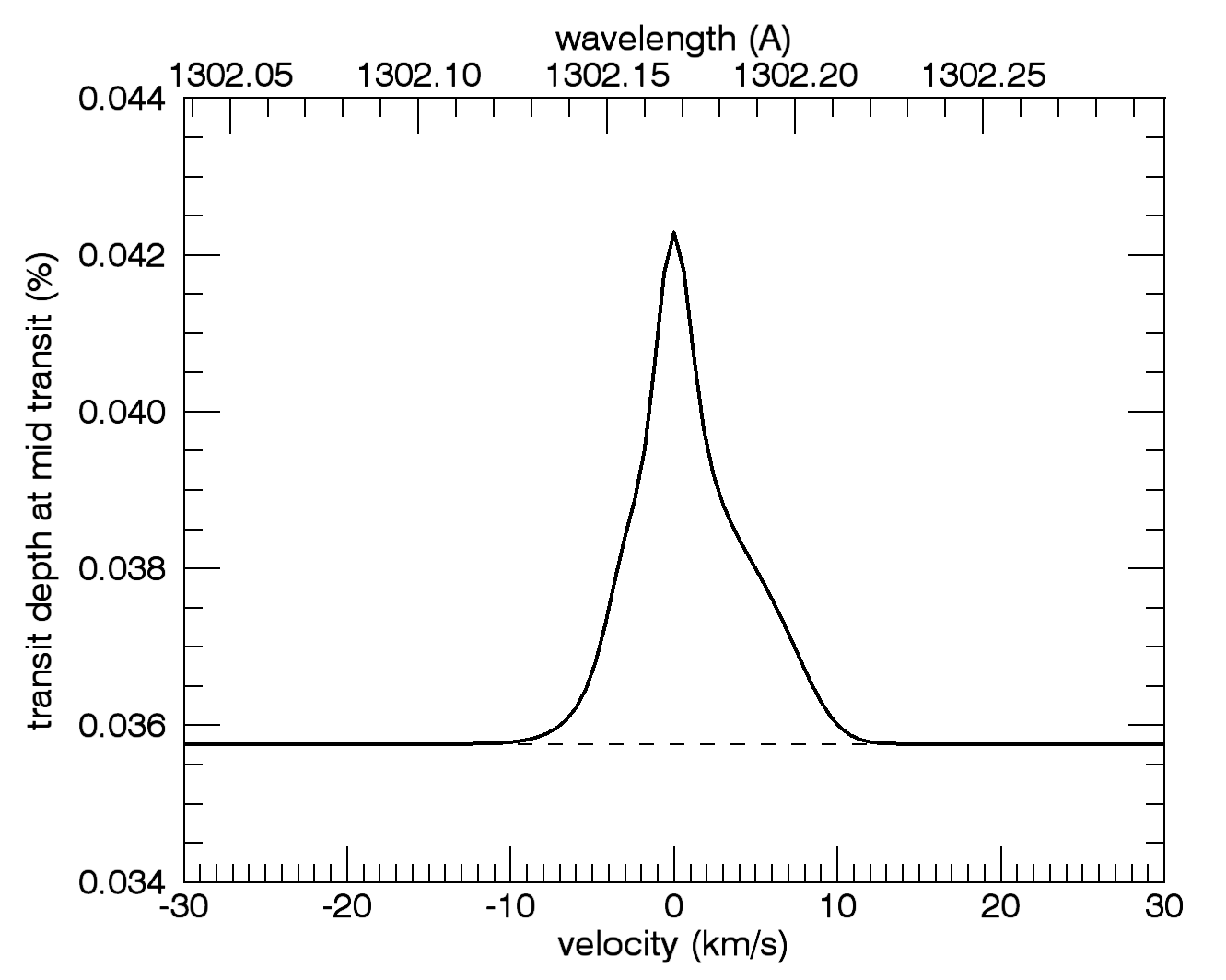}
\caption{Transit depth as a function of velocity (wavelength) in the O{\sc i} line at $1302.2$~\AA, assuming that the planet is at mid-transit. The dashed line indicated the transit depth of the broad-band optical transit.}
\label{fig.depth}
\end{figure}

Figure \ref{fig.depth} shows how the transit depth varies as a function of velocity and wavelength in the O{\sc i} line at $1302.2$~\AA, assuming that the planet is at mid-transit. The maximum transit depth is only $0.042\%$ near line centre. Out of Doppler shifts of $\pm 10$~km/s, the transit depth is essentially that of the broad band transit ($0.036\%$, dashed line). The reason for such a small increase in transit depth in the O{\sc i} $1302.2$~\AA\ line is that the atmosphere remains mostly optically thin in this line, with $\tau_v$ reaching maximum values of $\sim 1$ near zero velocities for regions very close to the planetary surface.  Such a small increase in transit depth in the O{\sc i} $1302.2$~\AA\ line is unlikely to be detectable. The lack of significant line broadening is due to the low velocity material (Figure \ref{fig.Vlos}) in the exosphere and its low temperature. The relatively low column density of O{\sc i} (ranging from $10^{10}$ to $10^{13}$~cm$^{-2}$) is responsible for the small increase in transit depth. For comparison, \citet{2013A&A...553A..52B} reported column densities of $8\e{15}$~cm$^{-2}$ in the exosphere of the hot Jupiter HD189733, which generates a transit depth of 3.5\% (the broad band transit depth is $\sim3\%$). Our maximum value of column densities are nearly 3 orders of magnitude smaller, resulting in very small attenuations caused by the planetary atmosphere.

%
\section{Discussion and conclusion}\label{sec.conclusions}
In this paper, we modelled the wind of \hd\ and used it to predict the surface sputter yields and the particle distribution of the {refractory-rich} exosphere for the rocky planets \hd\,b and c. Our stellar wind model is possibly the most well constrained to date after that of the Sun. We used observationally-derived maps of stellar surface magnetic field \citepalias{paper1} for the inner boundary of our 3D wind model. Additionally, the mass-loss rate  derived in our wind model ($1.6\e{-14}\msano$) is constrained by Ly-$\alpha$ observations of the stellar astrosphere \citepalias{paper1}. We then used the results of our stellar wind model to quantify wind-induced surface sputtering for the two innermost rocky planets. With that, we were able to estimate the density and structure of the planetary exospheres, on the assumption that both planets do not have strong magnetic fields and have lost both primary (hydrogen-dominated) and secondary (CO$_2$-dominated) atmospheres through escape processes driven by the high-energy stellar flux.

Our results can be summarised as follows. The large-scale magnetic field of the planet-hosting star \hd\ can be described as a dipole whose axis is roughly perpendicular to the stellar rotation axis. As a consequence, the stellar wind of \hd\ is highly non-axisymmetric, which implies that planets orbiting in the equatorial plane of the star interact with low and high speed winds in a very short timescale. For example, at every planetary year (3-days orbit), planet `b' interacts with the local stellar wind whose velocities vary from $200$ to $315$~km~s$^{-1}$. A similar level of variation is also seen by planet `c' along its orbit of roughly one week.

Because of the close orbital distances, the stellar wind conditions around \hd\,b and c are much harsher than, for example, the solar wind conditions around the Earth, or even around Mercury. If these exoplanets were to have a magnetic field similar to that of the Earth, their magnetospheres would extend out to about 4 planetary radii (only one third of the Earth's magnetospheric size). However, if these planets were to have a magnetic field similar to that of Mercury, their magnetospheres would be crushed into the planets' surface and the stellar wind would directly interact with the planets' crust. In the latter case, due to the close proximity of both planets to the host star, the high {flux of particles} of the incident stellar wind makes the surfaces of the planets to sputter, similarly to what occurs on Mercury.

Based on a three-dimensional sputtering model created for Mercury \citep{2015P&SS..115...90P}, we simulated here sputtering processes induced by the stellar wind on the rocky planets \hd\,b and c. Our simulations showed that sputtering processes release refractory elements from the entire dayside surface with velocities sufficiently high to allow for elongated trajectories of the sputtered particles. In particular, we find that oxygen and magnesium are expected to form an extended neutral exosphere with densities larger than 10\,cm$^{-3}$, within several planetary radii. Because of the close proximity of both planets to the host star, a substantial amount of the neutral atoms will quickly be ionised and picked up by the stellar wind. Our simulations suggest the column density of O{\sc i} to be as large as $\sim$10$^{13}$~cm$^{-2}$ close to the day-side of planet `b' and a few times smaller and less extended for planet `c'.

The column densities are not symmetric, with enhanced densities ahead of the planets' orbits. This happens due to the angle that the velocity vector of the stellar wind particles makes with the day-side of the planet, when accounting for the orbital motion of the planet through the stellar wind. This enhanced column density ahead of the planet motion could cause an asymmetric transit, with a longer ingress phase than the egress phase, should it be observable. To assess its observability, we used a 3D ray tracing technique to calculate the transit depth in the O{\sc i} $1302.2$~\AA, showing that it is at most $0.042\%$ near line centre, i.e., only a small increase compared to the transit depth in the optical ($0.036\%$).  Such a small increase in transit depth in the O{\sc i} $1302.2$~\AA\ line is unlikely to be observable with current UV instrumentation.

\section*{Acknowledgements}
The authors wish to acknowledge the SFI/HEA Irish Centre for High-End Computing (ICHEC) for the provision of computational facilities and support. This work used the BATS-R-US tools developed at the University of Michigan Center for Space Environment Modeling and made available through the NASA Community Coordinated Modeling Center. 
AAV and CF acknowledge joint funding received from the Irish Research Council and Campus France through the Ulysses funding scheme. LF acknowledges useful discussions with Lena Noack and Ildar Shaikhislamov. GV thanks the Russian Science Foundation (project No. 14-50-00043, ``Exoplanets'') for support of his participation in international studies of exoplanets with photometric and spectroscopic observations. This work is based on observations made with the NASA/ESA Hubble Space Telescope, obtained from MAST at the Space Telescope Science Institute, which is operated by the Association of Universities for Research in Astronomy, Inc., under NASA contract NAS 5-26555. These observations are associated with program No. 14461.

\let\mnrasl=\mnras

\bsp
\label{lastpage}

\appendix
\section{Effects of gravitational force, rotation and photoionisation on the exospheric densities}\label{ap.physical}
\begin{figure*}
\includegraphics[width=2\columnwidth]{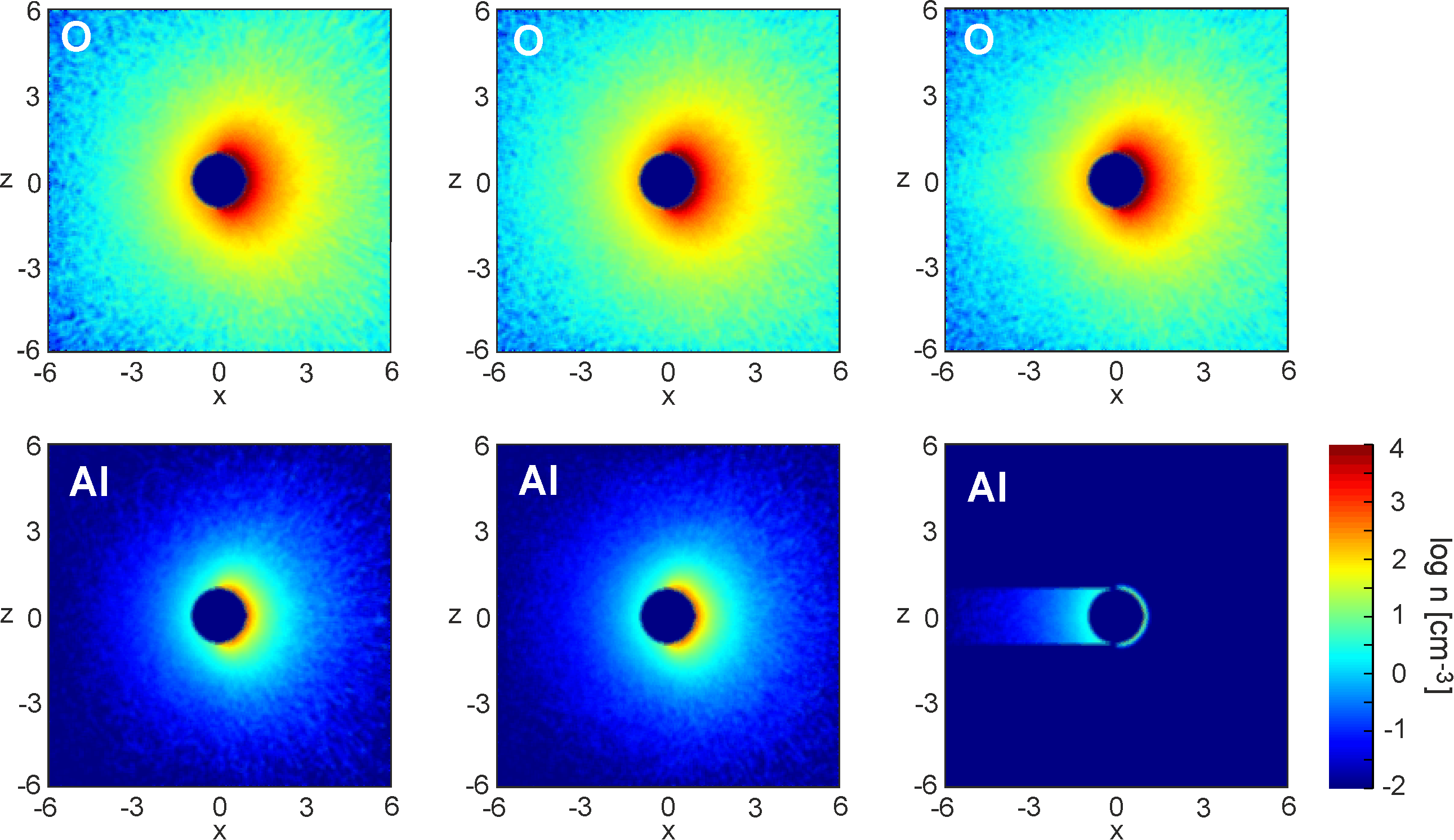}
\caption{Number density for sputtered oxygen and aluminium in the noon-midnight plane of planet `b'. The left panels are obtained by neglecting the influence of the stellar gravity, of the spin of the planet, and of photoionisation. The middle panel includes the stellar gravity and the planetary spin and orbital motion. The right panels are like the middle panels, but also accounting for photoionisation. The $x$-axis is in the planet-star direction, with the star on the right and the $z$-axis is perpendicular to the orbital plane. The scaling of the axes is given in planetary radii.}
\label{fig.dens_O_Al}
\end{figure*}

Figure~\ref{fig.dens_O_Al} illustrates the number density of sputtered oxygen and aluminium atoms for planet `b' in the noon-midnight plane. The different panels show the modelling results under different physical conditions and assumptions: no stellar gravitational attraction, no centrifugal force, and no photoionisation (left), no photoionisation (middle), all physical phenomena included (right). This is to highlight the effects of the different phenomena taken into account. Although the individual trajectories of the sputtered particles are altered by the centrifugal force and the gravitational force of the star, the bulk density of the exosphere is little affected. Photoionisation, however, can substantially diminish the neutral density, as can be seen by inspection of the Al density. In the simulation, we assume no photoionisation in the geometrical shadow of the planet, which leads to a density enhancement of neutral particles in the planetary shadow. Therefore, this sharp transition, is an artefact of the model and will be smoother in reality, since photons will also be scattered into the shadow region.

\end{document}